\documentclass[a4paper,11pt]{article}
\pdfoutput=1 

\usepackage{jinstpub} 
\usepackage{siunitx}
\usepackage{amsmath}
\usepackage{verbatim}
\usepackage{overpic}
\usepackage{url}
\usepackage{xspace}
\usepackage{subfigure}
\usepackage[symbol]{footmisc}

\usepackage{hyperref}
\hypersetup{colorlinks,breaklinks,
  linkcolor=black,citecolor=blue,
  filecolor=blue,urlcolor=blue,
  pdfpagemode=UseNone
}

\usepackage{caption}

\usepackage{chngcntr}
\counterwithout{equation}{section}

\graphicspath{{./}{figure/}} 

\newcommand*{\fig}[1]{Figure~\ref{fig:#1}}

\newcommand*{\eq}[1]{eq.~(\ref{eq:#1})}

\usepackage{lineno}

\title{\boldmath Study of MALTA2, a Depleted Monolithic Active Pixel Sensor, with grazing angles at CERN SPS 180\,GeV/c hadron beam}

\author[a,*]{L. Li,\note{Corresponding author.}}
\author[a]{P. Allport,}
\author[b]{I. Asensi Tortajada,}
\author[c]{P. Behera,}
\author[d,e]{D. V. Berlea,}
\author[f]{D. Bortoletto,}
\author[g]{C. Buttar,}
\author[b]{V. Dao,}
\author[c]{G. Dash,}
\author[d,e]{L. Fasselt,}
\author[b]{L. Flores Sanz de Acedo,}
\author[f]{M. Gazi,}
\author[a,\dagger]{L. Gonella, \note{Now at University of Trieste, Via Valerio 2, 34127, Trieste, Italy}}
\author[h]{V. Gonzalez,}
\author[b]{G. Gustavino,}
\author[b]{S. Haberl,}
\author[b]{T. Inada,}
\author[c]{P. Jana,}
\author[b]{H. Pernegger,}
\author[b]{P. Riedler,}
\author[b]{W. Snoeys,}
\author[b]{C. A Solans Sanchez,}
\author[b]{M. van Rijnbach,}
\author[b,h]{M. Vazquez Nunez,}
\author[c]{A. Vijay,}
\author[d,i]{J. Weick,}
\author[d]{S. Worm}

\affiliation[a]{University of Birmingham, Edgbaston Park Rd, B15 2TT, Birmingham, United Kingdom}
\affiliation[b]{CERN, Esplanade des Particules 1, 1211, Meyrin, Switzerland}
\affiliation[c]{Indian Institute Technology Madras, Hostel Ave, Tamil Nadu 600036, Chennai, India}
\affiliation[d]{Deutsches Elektronen-Synchrotron DESY, Platanenallee 6, 15738, Zeuthen, Germany}
\affiliation[e]{Institut f\"{u}r Physik, Humboldt-Universit\"{a}t zu Berlin, Newtonstrasse 15, 12489, Berlin, Germany}
\affiliation[f]{University of Oxford, Keble Road, OX1 3RH, Oxford, United Kingdom}
\affiliation[g]{University of Glasgow, University Ave, G12 8QQ, Glasgow, United Kingdom}
\affiliation[h]{Universitat de Val\'{e}ncia, Avinguda Blasco Ib\`{a}nez, 13, 46010, Valencia, Spain}
\affiliation[i]{TU Darmstadt, Karolinenplatz, 5, 64289, Darmstadt, Germany}

\emailAdd{long.l@cern.ch}

\abstract{MALTA2 is a Depleted Monolithic Active Pixel Sensor designed to meet the challenging requirements of future collider experiments, 
in particularly extreme radiation tolerance and high hit rate. The sensor is fabricated in a modified Tower 180 nm CMOS imaging technology to mitigate 
performance degradation caused by 100 MRad of Total Ionising Dose and greater than $10^{15}$ 1\,MeV $\text{n}_\text{eq}$/$\text{cm}^2$ 
of Non-Ionising Energy Loss. MALTA2 samples have been tested during the CERN SPS test beam campaign in 2023-2024, before and after irradiation at a fluence 
of 1\,$\times$\,$10^{15}$ 1\,MeV $\text{n}_\text{eq}$/$\text{cm}^2$.  The sensors were positioned at various inclinations relative to the beam, covering 
grazing angles from 0 to 60 degrees. This contribution presents measurements of detection efficiency and cluster size as functions of these angles, along 
with an estimation of the active depth of the depleted region based on the test beam results.}


\proceeding{11$^{\text{th}}$ International Workshop on Semiconductor Pixel Detectors for Particles and Imaging\\
  18-22 Nov 2024\\
  Strasbourg}

\begin{document}
\maketitle
\flushbottom

\section{Introduction}
\label{sect:intro}

Monolithic pixel detectors are widely considered for applications in high-energy physics experiments \cite{starPXL, alpide}, aiming to meet increasingly stringent 
requirements for vertexing and tracking detectors. Compared to hybrid pixel detectors, monolithic pixel detectors offer advantages in tracking 
applications due to their high granularity, low material budget, and low power consumption. Additionally, their fabrications using standard CMOS production 
processes from commercial foundries ensure lower costs and higher production yields. However, their radiation tolerance in dense collision environments remains 
a critical area of investigations. 

The MALTA series of sensors, based on Depleted Monolithic Active Pixel Sensor (DMAPS), are designed to fulfill the requirements for the future collider experiments. 
They initially target radiation tolerance levels of 100\,MRad for Total Ionizing Does (TID) and $>$\,$10^{15}$ 1\,MeV $\text{n}_\text{eq}$/$\text{cm}^2$ for 
Non-Ionising Energy Loss (NIEL), which are required by the outer pixel barrel of ATLAS ITk \cite{ITK_TDR} . A critical challenge for these sensors is mitigating 
signal degradation caused by a reduced active region and charge trapping under such high fluence conditions. To investigate these radiation-induced effects on 
DMAPS pixels, a grazing angle technique \cite{GrazingAngleTechnique} was performed on MALTA2 samples both before and after irradiation at a fluence of 
1\,$\times$\,$10^{15}$ 1\,MeV $\text{n}_\text{eq}$/$\text{cm}^2$, using the CERN Super Proton Synchrotron (SPS) 180\,GeV/c hadron beam. The detection 
efficiency and cluster size as functions of the inclined angles are presented, along with an estimation of the active depth of the depleted region for 
an irradiated MALTA2 sensor.

\section{MALTA2}
\label{sect:MALTA2}
MALTA2 is the second-generation sensor of the MALTA series, featuring optimizations to reduce Random Telegraph Signal (RTS) noise, thereby enabling an operational threshold 
of approximately 100\,$\text{e}^-$ \cite{Mini-MALTA, MALTA2_FE}. The sensor is fabricated using a Tower 180\,nm CMOS imaging process, modified with an additional 
low dose N-type implant \cite{MALTA_process_modification}. It  
features a matrix of 512\,$\times$\,224 pixels within a total area of 1.8\,$\times$\,0.9\,$\text{cm}^2$. The pixel pitch is 36.4\,$\times$\,36.4\,$\text{cm}^2$, and the 
sensor thickness is available in 50\,$\mu$m, 100\,$\mu$m and 300\,$\mu$m. The sensors are produced on two types of high-resistivity ($>1\,\text{k}\Omega\text{cm}$) p-type substrate: an epitaxial 
layer (EPI) and a Czochralski (Cz) bulk. The latter is optional for larger reverse substrate voltages ($\text{V}_\text{sub}$), yielding a larger depleted region and better radiation tolerance \cite{MALTACz}. 
MALTA2 employs a small collection electrode design measuring 3\,$\times$\,3\,$\mu\text{m}^2$, resulting in minimal capacitance (< 5\,fF), low noise level
(< 5\,$\text{e}^-$) and low power consumption ($\sim$\,1\,$\mu\text{W}/\text{pixel}$). 
To enhance the charge collection when the signal charge is generated at the pixel 
corners, two additional process modifications have been introduced in MALTA2: a gap in the low dose N-type implant (NGAP) at the pixel edge or an extra P-type implant at 
the same location (XDPW). Both modifiactions improve the charge collection and thus the detection efficiency \cite{MALTA_XDPW_NGAP}. 
The chip features an asynchronous read-out scheme that avoids 
propagating a clock signal throughout the pixel matrix, further reducing the power consumption from the digital part. 
The hit information is transmitted from the in-pixel circuit to the periphery, according to a sequence of short pulses, when a particle transverses the sensor. 
In this work, Cz sensors with a thickness of 100\,$\mu$m and XDPW modification were characterized with test beam for performance analysis.

\section{The MALTA telescope}
\label{sect:telescope}
The MALTA telescope \cite{MALTA_telescope}, developed for characterizing pixel detector prototypes, is permanently installed at H6 beam line in the North Area of SPS at CERN. 
During the MALTA2 test beam campaign in 2023-2024, mixed hadron beams with a momentum of approximately 180\,GeV/c were delivered. 

The telescope comprises six reference planes employing MALTA chips. A scintillator, placed behind reference planes, provides a precise time 
reference. The trigger system is fully configurable, which enables triggering on coincidence between the scintillator and the telescope planes. A cold box, which can 
contain up to two Devices Under Test (DUT), is positioned between the innermost reference planes. The corresponding cooling system provides 
a temperature down to -15\,$^\circ$C and keeps relative humidity below $1\%$. In addition, a rotational stage is responsible for the inclination 
of the DUT from -60$^{\circ}$ to 60$^{\circ}$ along the vertical axis with respect to the beam line. The spatial resolution of the telescope, 
extracted from the residual between the tracking intercept and the fastest hit of the closest cluster reconstructed of a DUT, 
is better than $5\,\mu\text{m}$.

\section{Grazing angle study}
\label{sect:GA}
Two MALTA2 chips, one unirradiated and the other irradiated to a fluence of 1\,$\times$\,$10^{15}$ 1\,MeV $\text{n}_\text{eq}/\text{cm}^2$, 
underwent the grazing angle technique. 
In the measurement, DUTs were rotated from 0$^{\circ}$ to 60$^{\circ}$
in 5$^{\circ}$ steps, relative to the particle track. 
\begin{figure}[htb!]
  \centering
  \subfigure[]{
    \includegraphics[width=0.48\textwidth]{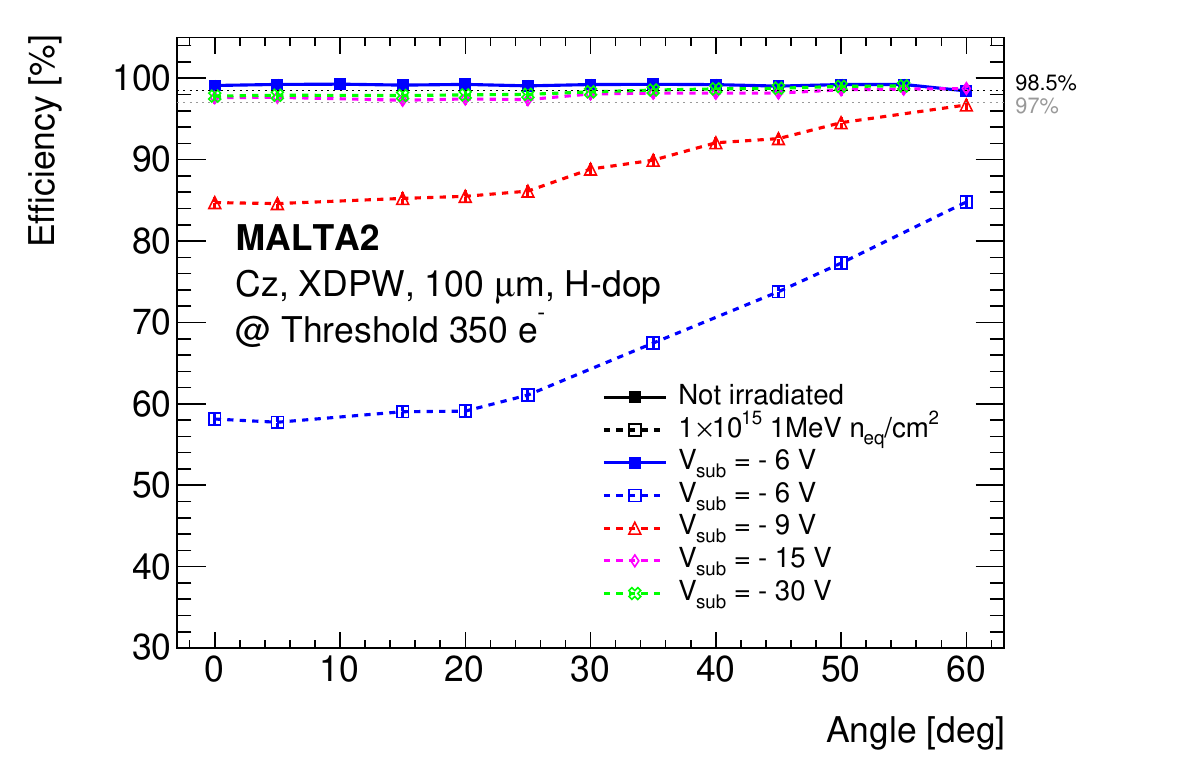}
    \label{fig:angle:eff}
  }
  \subfigure[]{
    \includegraphics[width=0.48\textwidth]{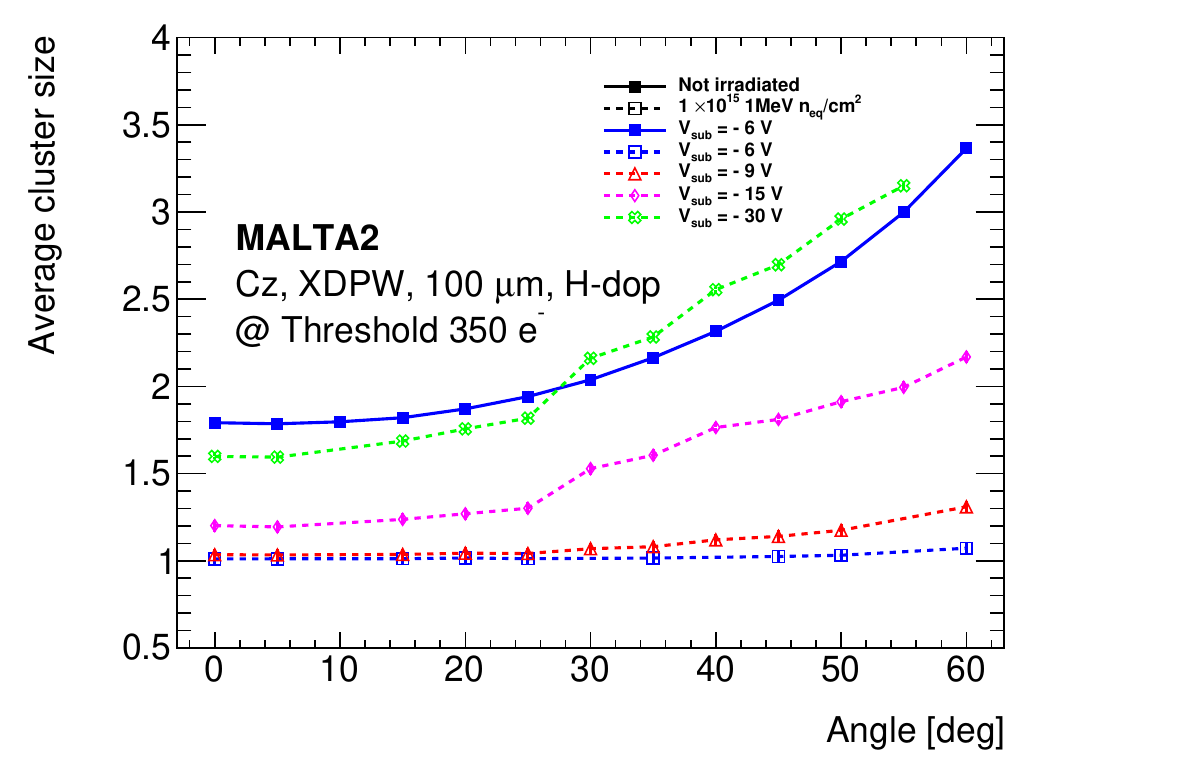}
    \label{fig:angle:clsize}
  }

  \caption{(a) Average efficiency and (b) average cluster size of irradiated and unirradiated MALTA2 samples
  as a function of incidence angle between the beam and the sensor. $\text{V}_\text{sub}$ is set at 6\,V 
  (6, 9, 15 and 30\,V for irradiated sample) and the p-well is reversely biasd at 6\,V.
  The operating threshold is around 350\,$\text{e}^-$. }
  \label{fig:angle}
\end{figure}
The average detection efficiency and the average cluster size, before and after irradiation,
as functions of rotational angles are depicted in \fig{angle}. Before irradiation, a full efficiency greater than $98.5\%$ was achieved, at $\text{V}_\text{sub}$ 
of 6\,V and with an operating threshold around 350\,$\text{e}^-$, regardless of the angles, indicating a sufficient active region and effective 
charge collection.
For the irradiated sample, a marked reduction in efficiency is observed at 0°, indicating charge collection loss due to a reduced active region and charge trapping. 
As the rotational angle exceeds 30°, the detection efficiency significantly recovers, attributed to the extended charge trajectory within the 
active region, enabling the collection of more signal charge. 
The average cluster size remains 1 across various rotational angles, showing suppressed 
charge sharing due to the reduced electric field at the pixel corners. To mitigate these degradations, the irrradiated sensor is operated at larger $\text{V}_\text{sub}$ 
(up to 30\,V). As shown in \fig{angle:eff}, the average efficiency significantly recovers with increasing $\text{V}_\text{sub}$, approaching a level comparable 
to that before irradiation when the voltage exceeds 15\,V. Similarly, the recovery of the average cluster size, shown in \fig{angle:clsize}, suggests an expanded active region 
and enhanced charge sharing due to an improved electric field at the pixel corners at larger $\text{V}_\text{sub}$.

\begin{figure}[htb!]
  \centering
  \includegraphics[width=0.45\textwidth, origin=c]{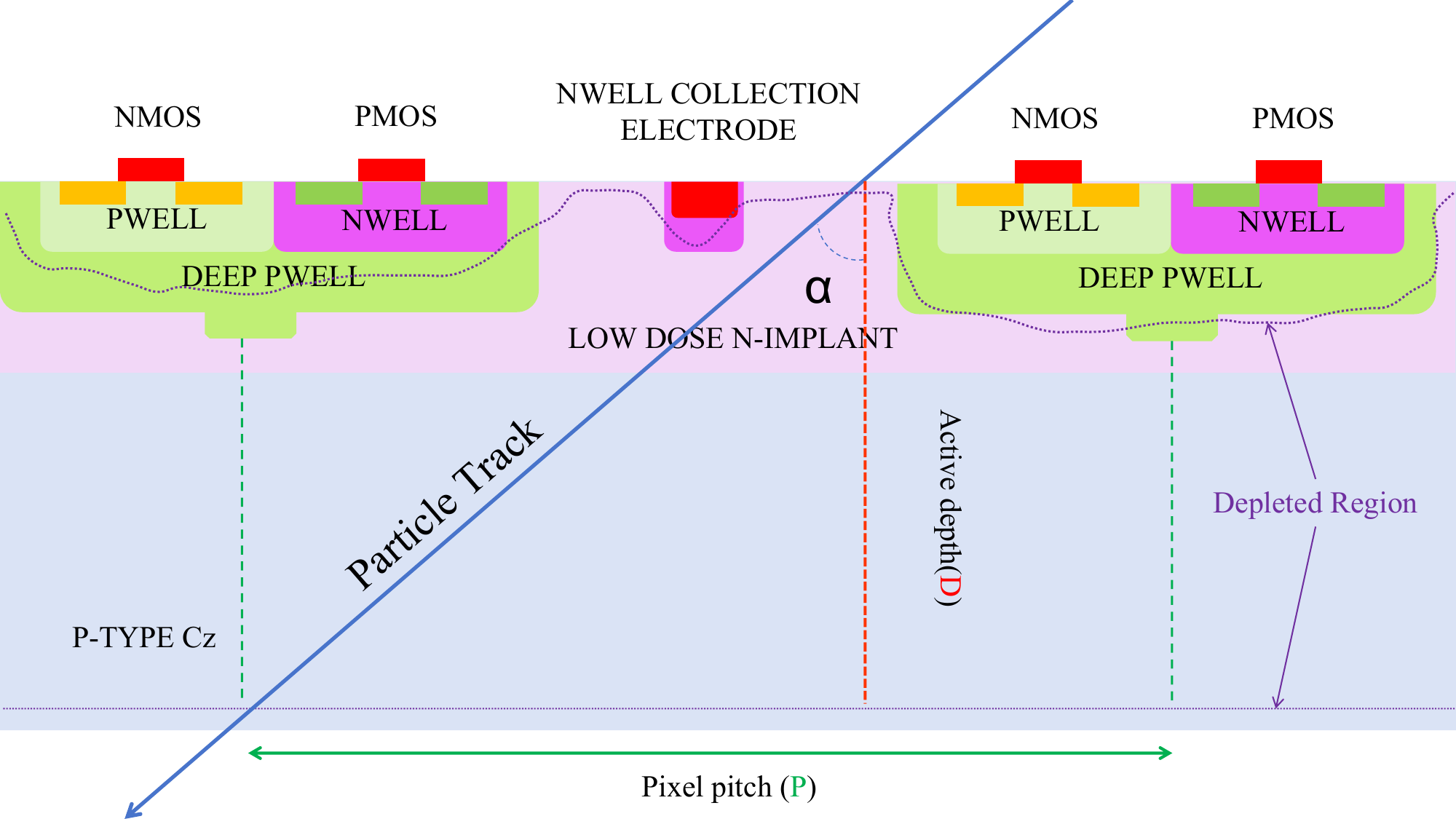}

  \caption{Illustration of the cross section of MALTA2 sensor and the grazing angle measurement method.}
  \label{fig:crossX}
\end{figure}

Additionally, the active depth, taking the irradiated sensor as an example, is estimated according to the grazing angle measurement results. 
As illustrated in \fig{crossX}, the geometrical relationship between the active depth and the cluster size can be described as
\begin{equation}
  \label{eq:grazing_geo}
  \text{Cluster size}_{\perp} (\alpha) = \frac{D}{P}\tan(\alpha) + \text{Cluster size}_{\perp} (0),
  \footnote[1]{This formula is purely geometrical, assuming ideal efficiency within the depleted region and no contribution from external regions.} 
\end{equation}  
where \textit{P} is the pixel pitch, $\alpha$ is the rotational angle, $\text{Cluster size}_{\perp} (0)$ and $\text{Cluster size}_{\perp} (\alpha)$ are the cluster sizes 
in the direction perpendicular to the rotational axis at angles of 0 and $\alpha$ respectively, and \textit{D} is the assumed active depth.
This method has been applied to measurements of unirradiated MALTA2 EPI sensors \cite{MALTA2_etct}, yielding an active depth comparable to that obtained from Edge Transient Current
Technique (E-TCT) measurements.

\begin{figure}[htb!]
  \centering
  \includegraphics[width=0.6\textwidth]{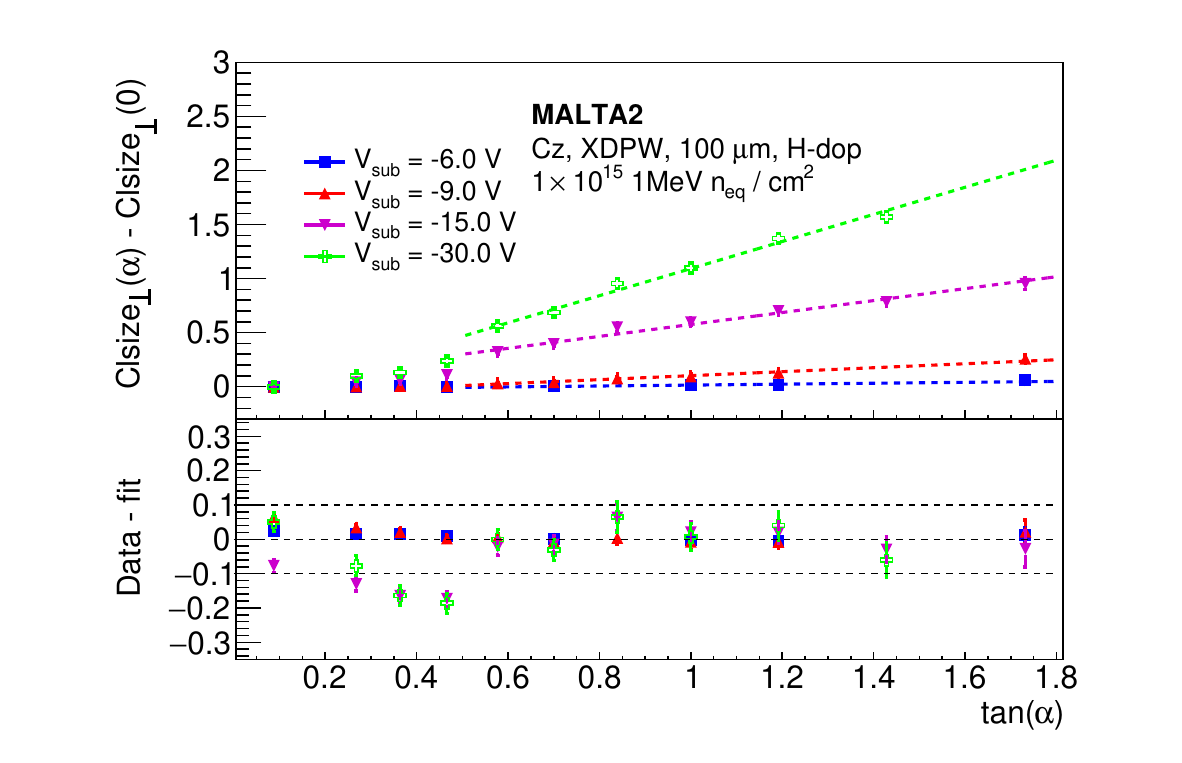}
  \caption{The linear fit of cluster $\text{size}_{\perp}$ residuals as a function of the tangent of the grazing angles. The sample is irradiated to 
  1\,$\times$\,$10^{15}$ 1MeV $\text{n}_\text{eq}$/$\text{cm}^2$ with $\text{V}_\text{sub}$ set at 6\,V, 9\,V, 15\,V and 30\,V, and an operating threshold
  around 350\,$\text{e}^-$.}
  \label{fig:linearfit}
\end{figure}

Shown in \fig{linearfit} are the residuals of the cluster size projected in the direction perpendicular to the rotational axis 
\footnote[1]{Cluster size in the direction parallel to the rotational axis stays constant} at angles $\alpha$ and 0 degrees,
plotted as functions of the tangent of $\alpha$. 
Excellent linearities are observed when $\tan (\alpha)$ exceeds 0.5. However, they break down at lower angular regions where charge diffusion dominates the cluster 
formation process. Thus, a linar fit is performed on each data set, and the active depth \textit{D} is extracted from the slope of the fitting function according to \eq{grazing_geo}.
This technique becomes ineffective at $\text{V}_\text{sub}$ of 6 V due to significant suppression of charge sharing between pixels. The estimated 
active depth increases with $\text{V}_\text{sub}$, and an approximation of 40\,$\mu$m is obtained at $\text{V}_\text{sub}$ of 30\,V.
The estimation is highly dependent on the threshold configuration, which significantly impacts the clustering process. Therefore, direct investigations via E-TCT are 
needed for further varifications.

\section{Conclusion}
MALTA2 is a full-scale DMAPS prototype of the MALTA project. Its detection performance, 
including detection efficiency and cluster size, before and after irradiation to a 
fluence of $1\,\times\,10^{15}\,\text{1\,MeV}\,\text{n}_\text{eq}/\text{cm}^2$, has been characterized with the grazing angle technique.  
Significant performance degradations were observed after irradiation, indicating a reduced active region and decreased electric field at the pixel corners. 
To mitigate these effects, the irradiated sensor is operated under larger $\text{V}_\text{sub}$, achieving an average detection efficiency of $97\%$, comparable to that
obtained before irradiation. 
In addition, an approxmation of 40\,$\mu$m for the active depth is obtained at $\text{V}_\text{sub}$ of 30\,V, 
with a threshold of around $350\,\text{e}^-$.

The conducted study provides important insights into sensor performance and radiation hardness, demonstrating the potential of MALTA2 for future collider experiment 
applications. Further investigations, through E-TCT, are planned to measure the active depth and verify the estimations made using the grazing angle technique.

\acknowledgments
This project has received funding from the European Union’s Horizon 2020 Research and Innovation programme under Grant Agreement numbers 101004761 (AIDAinnova), 
675587 (STREAM), and 654168 (IJS, Ljubljana, Slovenia).

\bibliographystyle{JHEP}
\bibliography{MALTA2}

\providecommand{\href}[2]{#2}\begingroup\raggedright\begin{thebibliography}{10}

\bibitem{starPXL}
G.~Contin et~al., \emph{{The STAR MAPS-based PiXeL detector}},
  \href{https://doi.org/10.1016/j.nima.2018.03.003}{\emph{Nucl. Instrum. Meth.
  A} {\bfseries 907} (2018) 60}
  [\href{https://arxiv.org/abs/1710.02176}{{\ttfamily 1710.02176}}].

\bibitem{alpide}
{\scshape ALICE} collaboration, \emph{{The ALPIDE pixel sensor chip for the
  upgrade of the ALICE Inner Tracking System}},
  \href{https://doi.org/10.1016/j.nima.2016.05.016}{\emph{Nucl. Instrum. Meth.
  A} {\bfseries 845} (2017) 583}.

\bibitem{ITK_TDR}
{\scshape ATLAS} collaboration, \emph{{Technical Design Report for the ATLAS
  Inner Tracker Strip Detector}},
  \href{https://doi.org/10.17181/CERN.FOZZ.ZP3Q}{\emph{CERN-LHCC-2017-021,
  ATLAS-TDR-025} (2017) }.

\bibitem{GrazingAngleTechnique}
S.~Meroli et~al., \emph{A grazing angle technique to measure the charge
  collection efficiency for {CMOS} active pixel sensors},
  \href{https://doi.org/https://doi.org/10.1016/j.nima.2010.12.122}{\emph{Nucl.
  Instrum. Meth. A} {\bfseries 650} (2011) 230}.

\bibitem{Mini-MALTA}
M.~Dyndal et~al., \emph{Mini-{MALTA}: radiation hard pixel designs for
  small-electrode monolithic {CMOS} sensors for the high luminosity {LHC}},
  \href{https://doi.org/10.1088/1748-0221/15/02/P02005}{\emph{Journal of
  Instrumentation} {\bfseries 15} (2020) P02005}.

\bibitem{MALTA2_FE}
F.~Piro et~al., \emph{{A 1$\mu$W Radiation-Hard Front-End in a 0.18-$\mu$m
  {CMOS} Process for the MALTA2 Monolithic Sensor}},
  \href{https://doi.org/10.1109/TNS.2022.3170729}{\emph{IEEE Trans. Nucl. Sci.}
  {\bfseries 69} (2022) 1299}.

\bibitem{MALTA_process_modification}
H.~Pernegger, \emph{Monolithic pixel development in towerjazz 180 nm {CMOS} for
  the outer pixel layers in the {ATLAS} experiment},
  \href{https://doi.org/https://doi.org/10.1016/j.nima.2018.07.043}{\emph{Nucl.
  Instrum. Meth. A} {\bfseries 924} (2019) 92}.

\bibitem{MALTACz}
H.~Pernegger et~al., \emph{{MALTA}-{C}z: a radiation hard full-size monolithic
  {CMOS} sensor with small electrodes on high-resistivity czochralski
  substrate},
  \href{https://doi.org/10.1088/1748-0221/18/09/P09018}{\emph{Journal of
  Instrumentation} {\bfseries 18} (2023) P09018}.

\bibitem{MALTA_XDPW_NGAP}
M.~Munker et~al., \emph{Simulations of {CMOS} pixel sensors with a small
  collection electrode, improved for a faster charge collection and increased
  radiation tolerance},
  \href{https://doi.org/10.1088/1748-0221/14/05/C05013}{\emph{Journal of
  Instrumentation} {\bfseries 14} (2019) C05013}.

\bibitem{MALTA_telescope}
M.~van Rijnbach et~al., \emph{{Performance of the {MALTA} telescope}},
  \href{https://doi.org/10.1140/epjc/s10052-023-11760-z}{\emph{Eur. Phys. J. C}
  {\bfseries 83} (2023) 581}
  [\href{https://arxiv.org/abs/2304.01104}{{\ttfamily 2304.01104}}].

\bibitem{MALTA2_etct}
D.V.~Berlea et~al., \emph{{Depletion depth studies with the {MALTA2} sensor, a
  depleted monolithic active pixel sensor}},
  \href{https://doi.org/10.1016/j.nima.2024.169262}{\emph{Nucl. Instrum. Meth.
  A} {\bfseries 1063} (2024) 169262}.

\end{thebibliography}\endgroup

\end{document}